\documentstyle[12pt,epsfig]{article}
\begin{document}

\noindent Stockholm\\
USITP 01\\
July 2001

\vspace{1cm}

\begin{center}

{\Large ${\bf CP}^n$, OR, ENTANGLEMENT ILLUSTRATED}

\vspace{1cm}

{\large Ingemar Bengtsson}\footnote{Email address: ingemar@physto.se. 
Supported by NFR.}

\

{\large Johan Br\"annlund}\footnote{Email address: jbr@physto.se.}

\

{\sl Stockholm University, SCFAB\\
Fysikum\\
S-106 91 Stockholm, Sweden}

\

{\large Karol \.Zyczkowski}\footnote{Email address: karol@cft.edu.pl}

\

{\sl Centrum Fizyki Teoretycznej, Polska Akademia Nauk\\
Al. Lotnikow 32/44, 02-668 Warszawa, Poland}

\vspace{8mm}

{\bf Abstract}

\end{center}

\vspace{5mm}

\noindent We show that many topological and geometrical properties of complex 
projective space can be understood just by looking at a suitably constructed 
picture. The idea is to view ${\bf CP}^n$ as a set of flat tori parametrized 
by the positive octant of a round sphere. We pay particular attention to 
submanifolds of constant entanglement in ${\bf CP}^3$ and give a few new 
results concerning them.

\newpage

{\bf 1. Introduction.}

\vspace{5mm}

\noindent Most people are familiar with the sphere ${\bf S}^2$; how to make 
a map of a sphere and how to use it to get insight into the 
geometrical and topological properties of a sphere. It is rare to see complex 
projective space ${\bf CP}^n$ treated in the same way, even though a large 
number of people deal with the geometrical and topological properties of this 
space in their everyday work. (It happens that ${\bf CP}^n$ is precisely the 
space of pure states of a particle of spin $n/2$, so indeed 
undergraduate physics students deal with it every day! For $n = 1$ it happens 
that ${\bf CP}^1 = {\bf S}^2$ and we get the ``Bloch sphere'' but for 
$n > 1$ ${\bf CP}^n$ is not a sphere.) 

It is the purpose of this paper to convince the reader---or at least those 
readers who like this sort of thing---that we can literally draw maps of 
${\bf CP}^2$ and ${\bf CP}^3$. Now every map distorts geography, so if we 
choose some property of ${\bf CP}^3$ that we wish to illustrate it is not 
obvious that it will be recognizable on the map. Our interest lies in 
illustrating the geometry of quantum mechanical entanglement, and the 
pleasant surprise is that our map works very well. 
Unfortunately we cannot really draw our map when the real dimension of 
the space to be mapped exceeds six, which means that we must restrict 
ourselves to the entanglement of two qubits in a pure state. This case is 
fully understood---it is probably the only case where this is true---so 
that we will not derive any new insights. Nevertheless we found the picture 
pleasing, and we would like to share it with others. 

The paper is organized as follows: In section 2 we introduce the picture 
and use it to illustrate some topological properties of ${\bf CP}^2$. In 
section 3 we study the separable and maximally entangled states in ${\bf CP}^3$. 
This is the lowest dimension where entanglement can occur. 
In section 4 we study submanifolds of states with intermediate 
entanglement. Section 5 is an aside on the Schmidt decomposition and the 
statistical geometry of density matrices and section 6 is another aside on 
the symplectic geometry of ${\bf CP}^n$. Throughout we try to keep track of 
which properties that are special to low dimensions, and which are not. Our 
main purpose is pedagogical although a few new results are included (eg. in 
sections 4 and 6). It will be helpful but we hope not necessary if the reader 
has a nodding acquaintance with the 3-sphere and the Hopf fibration; there 
are references \cite{Urbantke} that explain all that is 
needed in elementary terms. It only remains to add that the picture was not 
invented by us. There is a branch of mathematics called toric geometry 
\cite{Ewald} whose subject matter (roughly speaking) consists of spaces that 
can be depicted in this way. Moreover the picture has featured in the quantum 
mechanics literature already \cite{Arvind} \cite{Nuno}.

\vspace{1cm}

{\bf 2. The picture.}

\vspace{5mm}

\noindent A pure state in quantum mechanics is described by a vector in 
an $N$ complex dimensional vector space; in Dirac's notation 

\begin{equation} |{\Psi}\rangle = \sum_{{\alpha} = 0}^n 
Z^{\alpha}|{\alpha}\rangle \ , \end{equation}

\noindent where $|{\alpha}\rangle $ is a given orthonormal basis, $n = 
N - 1$ and $Z^{\alpha}$ has $N$ complex components. It is 
understood that vectors that differ only by an overall complex factor count 
as the same state. This means that there is a one-to-one correspondence 
between the set of pure states and the set of equivalence classes

\begin{equation} (Z^0, Z^1, Z^2, \ ... , Z^n) \sim z 
(Z^0, Z^1, Z^2, \ ... , Z^n) \ , \hspace{5mm} z \in {\bf C}  \end{equation}

\noindent By definition this is complex projective space ${\bf CP}^n$. The 
numbers $Z^{\alpha}$ are known as homogeneous coordinates. To make 
this description more concrete, suppose $n = 2$. Choose the complex number 
$z$ and the relative phases ${\nu}_1$ and ${\nu}_2$ so that 

\begin{equation} (Z^0, Z^1, Z^2) = (n_0, n_1e^{i{\nu}_1}, n_2e^{i{\nu}_2}) 
\ , \end{equation}

\noindent where $0 \leq {\nu}_i < 2{\pi}$ and the real numbers $n_0, n_1, n_2$ 
are non-negative, $n_0 \geq 0 \ , n_1 \geq 0 \ , n_2 \geq 0$, and obey the 
constraint 

\begin{equation} n_0^2 + n_1^2 + n_2^2 = 1 \ . \label{sfar} \end{equation}

\noindent (So we work with normalized state vectors.) This means that the 
set of allowed numbers $n_0, n_1, n_2$ are in one-to-one correspondence 
with points on the positive octant of the 2-sphere, while the periodic 
coordinates ${\nu}_1, {\nu}_2$ are in one-to-one correspondence with 
the points on a torus. The description breaks down at the edges of the 
octant since then the torus is undefined. Apart from this we already have a 
picture of the topology of ${\bf CP}^2$ which turns out to be quite useful. 

Even more to the point, this picture reflects the geometry of ${\bf CP}^2$. 
There is a natural way to define the "distance" between 
two pure states in quantum mechanics---namely in the sense of statistical 
distance \cite{Wootters}. There is also a mathematically natural notion of 
distance on ${\bf CP}^n$ called the Fubini-Study metric. The 
two of them coincide; the distance $d$ between two states is given by 

\begin{equation} \cos^2{d} = \frac{|\langle \Psi_1 | {\Psi}_2 \rangle|^2}
{\langle \Psi_1 | {\Psi}_1\rangle \langle {\Psi}_2 |{\Psi}_2 \rangle } 
= \frac{|{Z}_1\cdot \bar{Z}_2|^2}{Z_1\cdot \bar{Z}_1 Z_2\cdot \bar{Z}_2} \ , 
\label{Study} \end{equation}

\noindent where $Z\cdot \bar{Z} = Z^{\alpha}\bar{Z}_{\alpha}$ and 
$\bar{Z}_{\alpha}$ is the row vector whose entries are the complex conjugates 
of the entries of the column vector $Z^{\alpha}$. Note that the maximum 
distance between two points equals ${\pi}/2$; the precise number is a convention 
but the fact that there exists a maximal distance is not. In infinitesimal form 
the Fubini-Study distance becomes the metric tensor 

\begin{equation} ds^2 = \frac{Z\cdot \bar{Z} dZ\cdot d\bar{Z} - Z\cdot d\bar{Z} 
dZ\cdot \bar{Z}}{Z\cdot \bar{Z} Z\cdot \bar{Z}} \ . \label{Fubini} \end{equation}

\noindent The point we are driving at is that this metric takes a very 
nice form in the coordinates given above; common algebraic work shows that 

\begin{eqnarray} ds^2 = dn_0^2 + dn_1^2 + dn_2^2 +  \hspace{5cm} \nonumber \\
\ \\ 
\hspace{3cm} + n_1^2(1 - n_1^2)d{\nu}_1^2 + n_2^2(1 - n_2^2)d{\nu}_2^2 - 
2n_1^2n_2^2d{\nu}_1d{\nu}_2 \ . \nonumber \label{metric} \end{eqnarray}

\noindent The first piece here, given eq. (\ref{sfar}), is recognizable 
as the ordinary "round" metric on the sphere. The second part is the metric 
on a flat torus, whose shape depends on where we are on the octant. Hence 
we are justified in thinking of ${\bf CP}^2$ as a set of flat 2-tori 
"parametrized" by a round octant of a 2-sphere. There is an evident 
generalization to all $n$. In particular for $n = 1$ we obtain a one 
parameter family of circles (that degenerate to points at the end of the 
interval). A moment's thought will convince the reader that this is 
simply a way to describe a 2-sphere---and indeed it is well known that 
${\bf CP}^1$ is the same thing as the 2-sphere ${\bf S}^2$, usually 
called the Bloch sphere when it is regarded as the space of states of a 
spin 1/2 particle. (The choice of the metric (\ref{Fubini}) means that 
the radius of this sphere is actually 1/2, so that the maximum distance 
between two points on the Bloch sphere is ${\pi}/2$.)

To make the case $n = 2$ quite clear we make a flat map of the octant. 
Two methods suggest themselves, namely stereographic \cite{Urbantke} and 
gnomonic projection. The 
first is a standard coordinate system with several advantages; notably one can 
cover an entire sphere minus one point with one map. For our purposes the 
gnomonic (central) projection works even better; here the projection from 
the sphere to the flat map is made from the center of the sphere. It does not 
matter that only half the sphere can be covered since we need to cover 
only one octant anyway. A decided advantage is that the geodesics on the sphere, 
i.e. its great circles, appear as straight lines on the map. This is obvious 
because a great circle is the intersection between the sphere and a plane 
through the origin, and this will appear as a straight line when we project 
from the origin. 

We choose to center 
the projection at the center of the octant and adjust the coordinate plane 
so that the coordinate distance between a pair of corners of the resulting 
triangle is one. Explicitly, first we choose an auxiliary basis in three space 
so that $X^0 = 1$ labels the center of the octant: 

\begin{equation} \begin{array}{lll} X^0 & = & \frac{1}{\sqrt{3}}(n_0 + 
n_1 + n_2) \\ X^1 & = & \frac{1}{\sqrt{2}}(- n_0 + n_1) \\ X^2 & = & 
\frac{1}{\sqrt{6}}(- n_0 - n_1 + 2n_2) \end{array} \ . \end{equation}

\noindent Next we define the gnomonic coordinates $x_1, x_2$ by 

\begin{equation} x_i = \frac{1}{\sqrt{6}}\frac{X^i}{X^0} \hspace{5mm} 
\Leftrightarrow \hspace{5mm} X^0 = \frac{1}{\sqrt{1 + r^2}} 
\hspace{5mm} X^i = \frac{\sqrt{6}x_i}{\sqrt{1 + 6r^2}} \ ; \hspace{4mm} r^2 
\equiv x_1^2 + x_2^2 \ . \end{equation}

\noindent The octant is bounded by great circles and therefore it now appears 
as a triangle centered at the origin. Its sides have coordinate length one 
and in the gnomonic coordinates the metric takes the form 

\begin{equation} ds^2 = dn_0^2 + dn_1^2 + dn_2^2 = \frac{6}{(1 + 6r^2)^2}
((1 + 6r^2)dx\cdot dx - 6(x\cdot dx)^2) \end{equation}

\noindent where $x\cdot dx \equiv x_1dx_1 + x_2dx_2$ and so on. Figure 1 
should make all this clear. 

\begin{figure}
        \centerline{ \hbox{
                \epsfig{figure=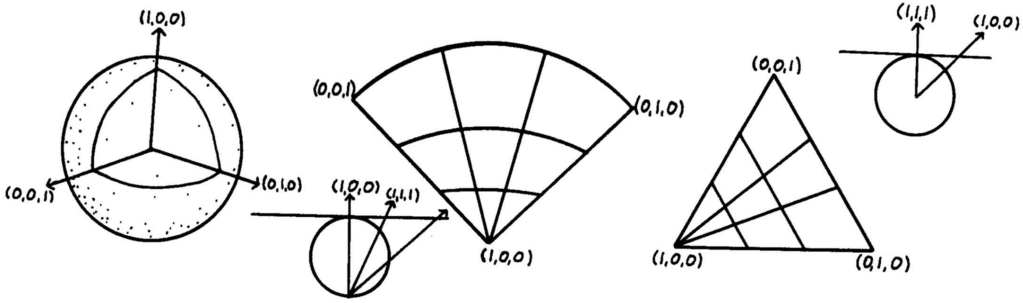,width=14cm}}}
        \caption{{\small An octant of a sphere and two flat maps: 
To the left we see the positive octant of the sphere. In the middle there 
is a stereographic map of the octant centered at $(n_0, n_1, n_2) = 
(1, 0, 0)$. On the right there is a gnomonic map centered at 
$\frac{1}{\sqrt{3}}(1,1,1)$. The projections are done respectively from 
the South Pole and from the center of the sphere, as shown. On each map 
four geodesics 
are drawn. A pair of these geodesics that meet an edge divides that edge 
into three equal parts. This gives an idea about how distances are 
distorted by the maps.}}
        \label{fig:ett}
\end{figure}

This takes care of the octant (the "manifold with corners" in the language of 
toric geometry \cite{Ewald}). It becomes a picture of ${\bf CP}^2$ when we 
remember that each point in the interior really represents a flat torus, 
conveniently regarded as a parallelogram with opposite sides identified. The 
shape of the parallelogram is relevant. According to eq. (\ref{metric}) the 
lengths of the sides are 

\begin{equation} L_1 = \int_0^{2{\pi}}ds = 2{\pi}n_1\sqrt{1 - n_1^2} 
\hspace{5mm} \mbox{and} \hspace{5mm} L_2 = 2{\pi}n_2\sqrt{1 - n_2^2} \ 
. \end{equation}

\noindent The angle between them is given by 

\begin{equation} \cos{{\theta}_{12}} = - \frac{n_1n_2}{\sqrt{1 - n_1^2}
\sqrt{1 - n_2^2}} \ . \end{equation}

\noindent The point is that the shape depends on where we are on the octant. 
So does the total area of the torus, 

\begin{equation} A = L_1L_2\sin{{\theta}_{12}} = 
4{\pi}^2n_0n_1n_2 \ . \end{equation}

\noindent The "biggest" torus occurs at the center of the octant. At the 
boundaries the area of the tori is zero. This is because there the tori 
degenerate to circles. Figure 2 shows how this happens. In effect an 
edge of the octant is a one parameter family of circles, in other words 
it is a ${\bf CP}^1$. 

\begin{figure}
        \centerline{ \hbox{
                \epsfig{figure=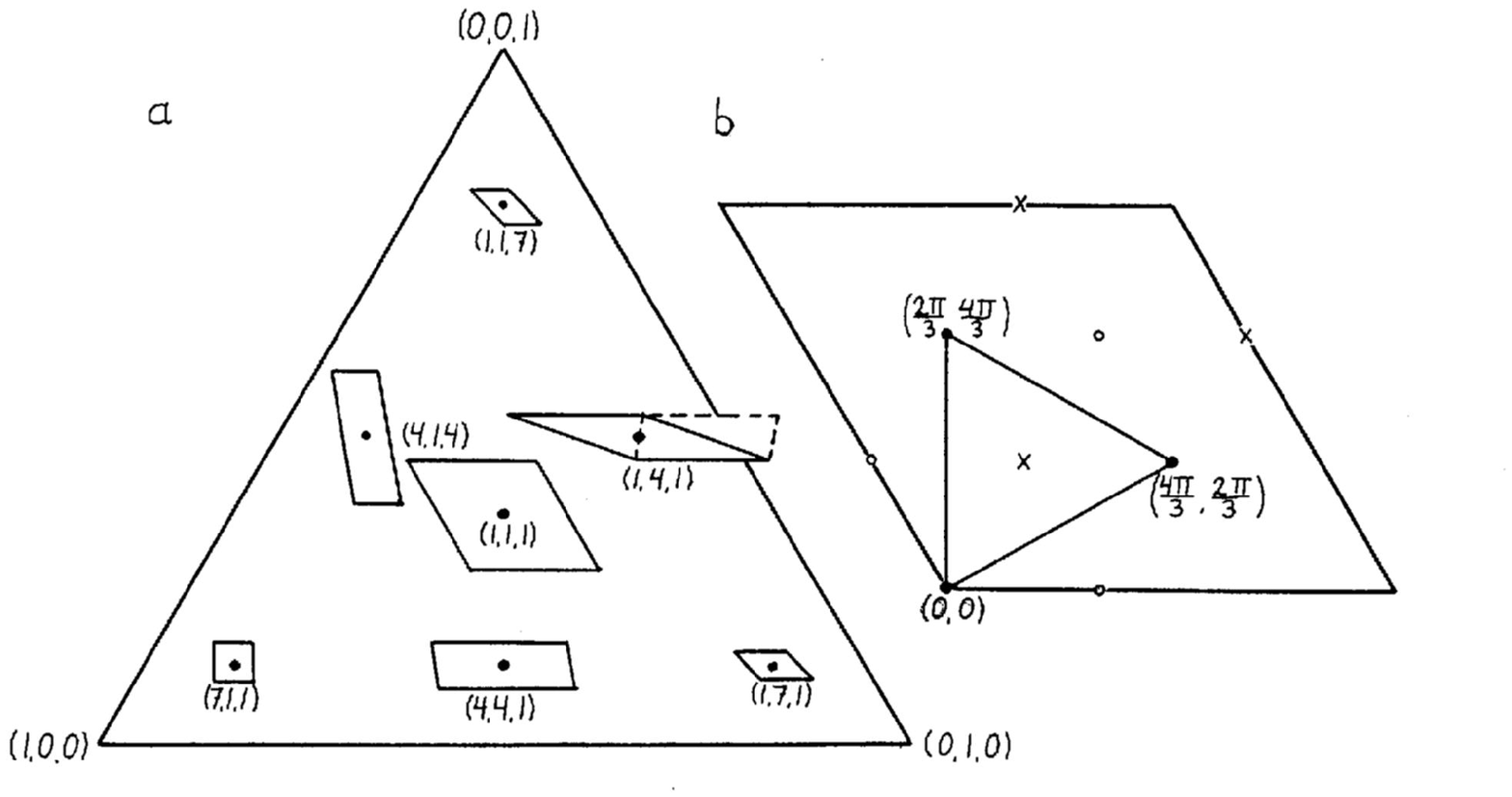,width=14cm}}}
        \caption{{\small The octant really parametrizes a family of flat 
tori of varying sizes and shapes. In a) we indicate how the torus lying over 
each interior point changes with position in the octant. The position in the 
octant is given by an unnormalized vector. At the edges the tori 
degenerate to circles so that the edges are complex projective lines, while 
the corners of the octant represent points. Sometimes it is convenient to perform 
a little cutting and gluing of the parallelogram before thinking about the shape 
of the torus it defines, as indicated with dashed lines for the torus lying over 
the point $(1,4,1)$. The size of the octant relative to that of the tori is 
exaggerated in the picture. To bring this home we show, in b), the largest 
torus---the one sitting over $(1,1,1)$---decorated with three times three 
points corresponding to three mutually unbiased bases (represented 
respectively by crosses and filled and unfilled dots). They are all 
mutually unbiased with respect to the basis 
that forms the corners of the octant. The coordinates $({\nu}_1, {\nu}_2)$ are 
given for one basis.}}
        \label{fig:tva}
\end{figure}

It is crucial to realize that there is nothing special going on at the 
edges and corners of the octant, whatever the impression left by the 
map may be. Like the sphere, ${\bf CP}^n$ is a homogeneous space and looks 
the same from every point. To see this, note that any choice of an orthogonal 
basis in a 3 dimensional Hilbert space gives rise to 3 points separated by 
the distance ${\pi}/2$ from each other in ${\bf CP}^2$. In projective geometry 
the triplet of points arising from an orthogonal basis in the underlying 
vector space is known as a triangle of reference. By an appropriate 
choice of coordinates we can make any such triple of points sit at the 
corners of an octant in a picture identical to the one above. 

To get used to the picture let us consider some submanifolds. 
${\bf CP}^2$ is also known as the "complex projective plane". Every pair of 
points in this "plane" defines a unique "complex projective line" containing 
the pair of points, and such a "line" is a ${\bf CP}^1$. Conversely a pair of 
complex projective lines always intersect at a unique point. Since a ${\bf CP}^1$ 
is always a sphere (of radius 1/2) the terminology may boggle some minds, 
but these intersection properties are precisely what defines a line in 
projective geometry. Through every point there passes a 2-sphere's worth 
of complex projective lines, conveniently parametrized by the way they intersect 
the "line at infinity", that is the set of points at maximal distance from 
the given point, which in itself is a ${\bf CP}^1$. This is easily illustrated 
provided we arrange the picture so that the given point sits in a corner. 

Another submanifold is the real projective plane ${\bf RP}^2$. It is defined 
in a way analogous to the definition of ${\bf CP}^2$ except that real rather 
than complex numbers are used. The points of ${\bf RP}^2$ are therefore in 
one-to-one correspondence with the set of lines through the origin in a 
three dimensional real vector space and also with the points of 
${\bf S}^2/{\bf Z}^2$, that is to say the sphere with antipodal points 
identified. In its turn this is a hemisphere with antipodal points on the 
equator identified. ${\bf RP}^2$ is clearly a subset of ${\bf CP}^2$. It is 
illuminating to see how the octant picture is obtained, starting from 
the stereographic projection of a hemisphere (a unit disk) and folding it 
twice, as in Figure 3. 

\begin{figure}
        \centerline{ \hbox{
                \epsfig{figure=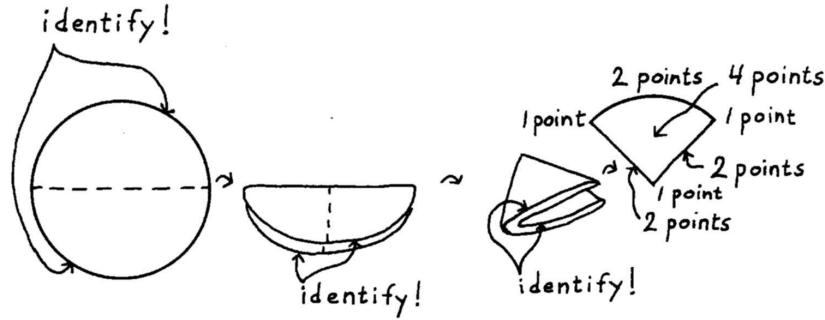,width=12cm}}}
        \caption{{\small Using stereographic rather than gnomonic 
coordinates we show how the 
octant picture of the real submanifold ${\bf RP}^2$ is related to the standard 
description as a hemisphere with antipodal points on the equator identified.}}
        \label{fig:tre}
\end{figure}

Brody and Hughston \cite{Brody} give a beautiful account of how the physics 
of a spin 1 particle is tied to the geometry of ${\bf CP}^2$. The space of 
all possible spin up states (with respect to some direction) 
forms a sphere of radius $1/\sqrt{2}$. This is not a complex projective 
line because of its size. The space of all possible spin 0 states (with 
respect to some direction) forms a sphere with antipodal points identified 
since such states do not distinguish up and down, hence this is an 
${\bf RP}^2$. This is illustrated in Figure 4 under the assumption that 
the corners of the octant correspond to eigenstates of the operator $S_z$. 
It is interesting to observe that if we start from a ${\bf CP}^1$ (and place it 
as an edge in the picture) then we can increase the size of the sphere 
by deforming the edge, but we cannot shrink it. Therefore ${\bf CP}^2$ contains 
non-contractible 2-spheres, just as a torus contains non-contractible 
circles. Note that two non-contractible 2-spheres always touch at a point, 
essentially because two complex projective lines intersect in a unique point. 

\begin{figure}
        \centerline{ \hbox{
                \epsfig{figure=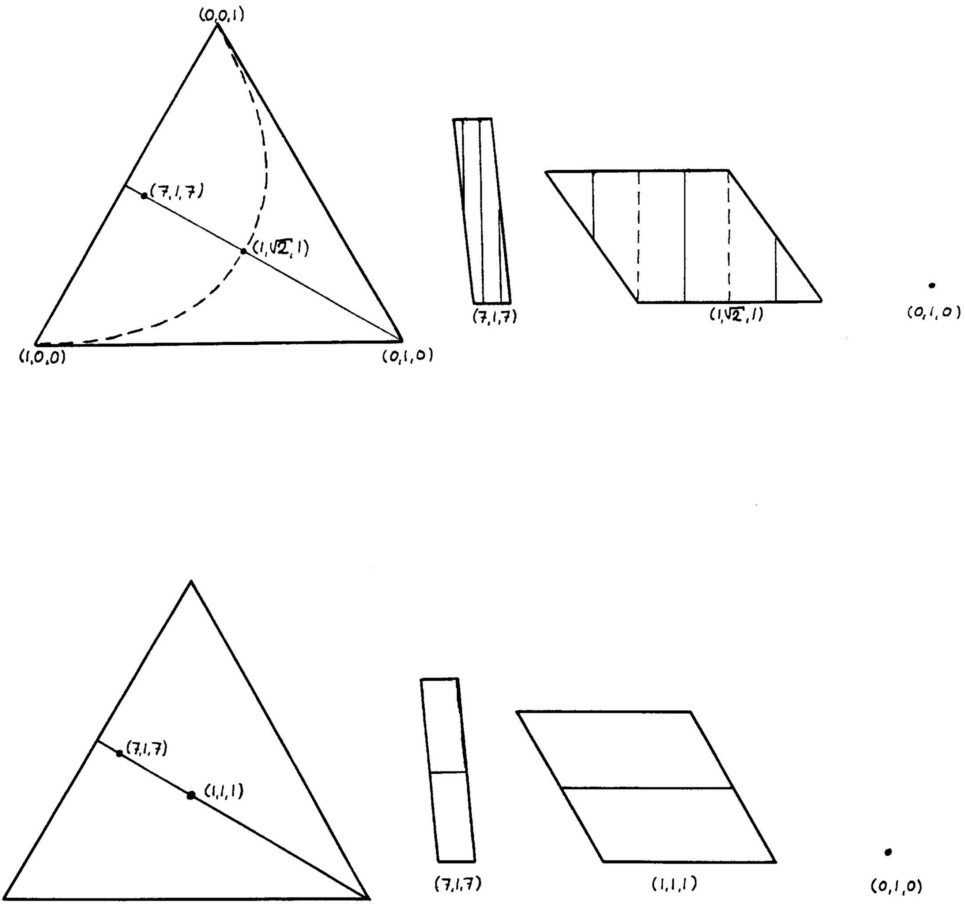,width=14cm}}}
        \caption{{\small The physics of spin systems (above). The dashed 
curve represents a sphere with radius 
$1/\sqrt{2}$, and the solid line an ${\bf RP}^2$. Both are one parameter 
families 
of circles with points at one end. The sphere has a point at both ends, while 
for the real projective plane the circle sits in the torus in such a way that 
when the torus is squashed to a circle then the circle wrapped inside it 
suddenly collapses to a circle of half the size it had just before. We show how 
this happens by drawing two of the tori explicitly; again points on the 
sphere are represented by dashed lines and points on the ${\bf RP}^2$ by 
solid lines. For comparison we also show a complex projective line (below). 
This is a one parameter family of circles with points at both ends, as shown 
in the tori on the right hand side. Note also that any edge of 
the octant is a ${\bf CP}^1$, namely the "line at infinity" with 
respect to the point in the opposite corner.}}
        \label{fig:fyra}
\end{figure}


Next choose a point. Place it at a corner of the octant and surround it with a 
3-sphere consisting of points at constant distance from the given point. 
In the picture this will appear as a curve in the octant with an entire torus 
sitting over each interior point of the curve. Readers familiar with 
the Hopf fibration know that ${\bf S}^3$ can be thought of as a one parameter 
family of tori with circles at the ends. The 3-sphere is round if the tori 
have a suitable rectangular shape. As we let the 3-sphere in Figure 5 grow 
the tori get more and more "squashed" by the curvature of ${\bf CP}^2$, and 
the roundness gradually disappears. When the radius reaches its maximum 
value of ${\pi}/2$ the 3-sphere collapses to a 
2-sphere, namely to the projective line "at infinity". Readers unfamiliar 
with the Hopf fibration may wish to consult Urbantke \cite{Urbantke} at 
this point; let us just remark that only odd dimensional spheres can 
be squashed in this specific way (so it is no use to try to picture this 
in terms of the 2-sphere). 

\begin{figure}
        \centerline{ \hbox{
                \epsfig{figure=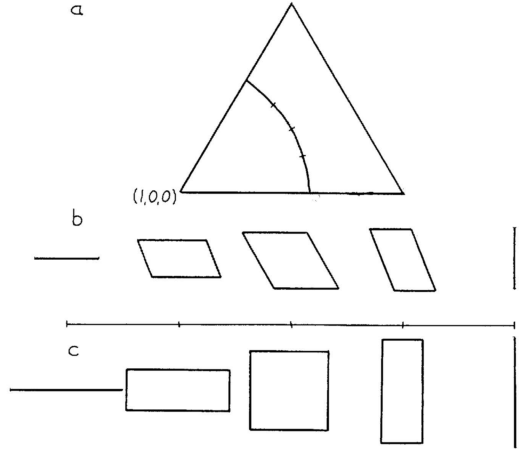,width=8cm}}}
        \caption{{\small The set of points at constant distance from a 
corner form a squashed 3-sphere. In a) we show how such a submanifold appears 
in the octant. All the points in the torus lying over a point on the curve are 
included. In b) we show how the size and shape of the torus change as we move 
along the curve in the octant; at the ends of the interval the tori 
collapse to circles. For comparison, in c) we show the corresponding picture 
for a round 3-sphere.}}
        \label{fig:fem}
\end{figure}


Finally, a warning: The picture distorts distances in many ways. For instance, 
the distance between two points in a given torus is shorter than it looks, 
because the shortest path between them is not the same as a straight line 
within the torus itself. Technically, the flat tori are not totally geodesic. 
Note also that we have chosen to exaggerate the size of the octant relative 
to that of the tori in the pictures. To realize how much room there is in the 
largest torus, consider mutually unbiased measurements \cite{Ivanovic}. It 
is known that one can find 4 sets of orthonormal bases in the Hilbert space 
of a spin 1 particle such that the absolute value of the scalar product 
between members of different basises is $1/\sqrt{3}$ ---clearly a kind of 
sphere packing 
problem if translated into geometrical terms by means of eq. (\ref{Study}). 
If one basis is represented by the corners of the octant, the remaining 3 times 
3 basis vectors are situated in the torus over the center of the octant. (See 
figure 2.)


\vspace{1cm}

{\bf 3. Separable and maximally entangled states in ${\bf CP}^3$.}

\vspace{5mm}

\noindent We now wish to illustrate entanglement. This forces us to increase 
the dimension, since our system should be composed of two subsystems. If we 
choose to study pairs of entangled qubits the complex Hilbert space is 
${\bf C}^2 
\otimes {\bf C}^2 = {\bf C}^4$, and the space of pure states becomes 
the six real dimensional space ${\bf CP}^3$. This is complex projective space. 
A preliminary remark about it is that it contains complex projective 
planes and lines (${\bf CP}^2$ and ${\bf CP}^1$) as submanifolds, and the 
intersection properties of these are just those of planes and lines in 
ordinary Euclidean space, with the important simplification that all 
troublesome exceptional cases (such as parallel planes that do not intersect 
in a line) are absent. 

It is easy to make a picture of this six dimensional space. Choose the 
coordinates 

\begin{equation} (Z^0, Z^1, Z^2, Z^3) = (n_0, n_1e^{i{\nu}_1}, n_2
e^{i{\nu}_2}, n_3e^{i{\nu}_3}) \ . \end{equation}

\noindent The phases now form a 3-torus (that we can picture as a rhomboid), 
while the non-negative real numbers $n_0$ etc. obey 

\begin{equation} n^2_0 + n_1^2 + n_2^2 + n_3^2 = 1 \ . \end{equation}

\noindent Hence they form a hyperoctant of the 3-sphere. Without further 
ado it is clear that if we perform a gnomonic projection centered at 

\begin{equation} (n_0, n_1, n_2, n_3) = \frac{1}{2}(1, 1, 1, 1) \end{equation}

\noindent then we obtain a picture of this hyperoctant as a regular 
tetrahedron. In this picture straight lines correspond to geodesics on the 
round hyperoctant. Over each point in the interior there is a flat 3-torus 
of a definite shape. The faces of the tetrahedron are complex projective 
planes, its edges are complex projective lines, and its corners are points. 
Such pictures will appear soon. 

We do not give the transformation to gnomonic coordinates here. This 
is because their only advantage is to give a nice symmetrical picture that is 
easy to draw, and most of our pictures can be drawn using only 
the fact that geodesics are straight lines, plus explicit knowledge of the 
two dimensional case. (Calculations should be performed in coordinate systems 
suited to calculation---for most purposes the embedding coordinates 
$n_0$, $n_1$, $n_2$, $n_3$ will do.) 

Now, what about entanglement? Our illustrations in this section will depict 
submanifolds of states of constant entanglement, namely separable states (that 
are not entangled) and maximally entangled states. In the next section we 
show how states of "intermediate entanglement" sit in ${\bf CP}^3$. The same 
story has been told before \cite{Kus} but not in quite this way. 

A brief review of the facts seems appropriate. State vectors for composite 
systems are conveniently written as 

\begin{equation} |{\Psi}\rangle = \frac{1}{\sqrt{N}} \sum_{i = 0}^n
\sum_{j = 0}^nC_{ij}|i\rangle |j \rangle \ , \label{17} \end{equation}

\noindent where $C_{ij}$ is an $N \times N$ matrix with complex entries and 
during the review we keep the state vector normalized. Throughout the 
discussion we rely on a fixed way of splitting the Hilbert space into a 
tensor product of two smaller Hilbert spaces. In other words it has been 
agreed that the Hilbert space is ${\cal H}_A\otimes {\cal H}_B$ in a 
specific way. Otherwise the term "entanglement" has no meaning. 
For the $2 \times 2$ case let us agree that 

\begin{equation} (Z^0, Z^1, Z^2, Z^3) \equiv (C_{00}, C_{01}, C_{10}, C_{11}) 
\ . \label{coordinates} \end{equation} 

\noindent The density matrix for the system can be written with composite 
indices, in the form 

\begin{equation} {\rho}_{ij, kl} = \frac{1}{N}C_{ij}C^*_{kl} \ . \end{equation}

\noindent It has rank one because the system is in a pure state. Now suppose 
that we are performing experiments on one of the subsystems only. Then the 
relevant density matrix is the partially traced density matrix ${\rho}_A = 
\mbox{Tr}_B{\rho}$, 

\begin{equation} {\rho}^A_{ik} = \sum_{j = 0}^n{\rho}_{ij,kj} \ . \end{equation}

\noindent The rank of this matrix may well be greater than one. There are 
two extreme cases. The global state of the system may be a product state, 

\begin{equation} |{\Psi}\rangle = |A\rangle |B\rangle = \sum_{i = 0}^n
\sum_{j = 0}^n (a_i|i\rangle )(b_j|j\rangle ) \hspace{5mm} \Leftrightarrow 
\hspace{5mm} C_{ij} = a_ib_j \ , \label{separable} \end{equation}

\noindent so the matrix $C$ is an outer (dyadic) product of two vectors $a$ and 
$b$. In this case the partially traced density matrix and the matrix 
$C_{ij}$ both have rank one and the individual subsystems are in pure states of 
their own. A global state of this kind is said to be un-entangled or 
{\it separable}. At the opposite end of the spectrum it happens that 

\begin{equation} {\rho}^A_{ik} = \frac{1}{N}{\bf 1} \hspace{5mm} \Leftrightarrow 
\hspace{5mm}  \sum_{j = 0}^nC_{ij}C^*_{kj} = {\delta}_{ik} \ . \label{SU} 
\end{equation}

\noindent This means that we know nothing at all about the state of the 
subsystems, 
even though the global state is precisely known. A global state of this kind 
is said to be {\it maximally entangled}. In between these cases are cases where 
the von Neumann entropy of ${\rho}^A$ takes some intermediate value. They are 
also entangled. It is known \cite{Everett} that an arbitrary state in the 
Hilbert space ${\cal H}^N \otimes {\cal H}^N$ can be brought to the form 

\begin{equation} |{\Psi}\rangle = \sum_{i = 0}^n c_i|i\rangle |i\rangle 
\label{Schmidt} \end{equation}

\noindent by means of unitary transformations belonging to the subgroup 
$U(N)\times U(N)$, that is by "local unitary" transformations acting 
independently on the two subsystems. The coefficients are square roots of 
the eigenvalues of the density matrix obtained when one subsystem is 
traced out, and hence they obey 

\begin{equation} \sum_{i = 0}^n c_i^2 = 1 \ . \label{Scmidtsimplex} 
\end{equation}

\noindent This is known as the Schmidt decomposition and is explained in 
many places \cite{Peres}\cite{Knight}. Once the Schmidt coefficients 
$c_i$ have been ordered 
(say by their sizes) they cannot be changed by local unitaries, so that they 
can be used to label the orbits of $U(N)\times U(N)$ in ${\bf CP}^3$. This ends 
our brief review of entanglement. 

Now we are going to take a look (literally!) on the submanifolds that we 
mentioned. Consider first the separable states, as described in eq. 
(\ref{separable}). If we 
keep $|A\rangle $ fixed and vary $|B \rangle $ we sweep out a ${\bf CP}^n$ that 
lies, in its entirety, in the submanifold. Reversing the roles of $|A\rangle $ 
and $|B\rangle $ we sweep out another ${\bf CP}^n$, and every state in the 
submanifold can be reached by a combination of these operations. This means 
that the submanifold of separable states is the Cartesian product ${\bf CP}^n 
\times {\bf CP}^n$. For $N = 2$ (that is $n = 1$) we get the four real 
dimensional submanifold ${\bf CP}^1\times {\bf CP}^1$ sitting inside 
${\bf CP}^3$, 
and the question is what it looks like in our picture. We want it to appear as a 
surface in the 3-torus, sitting over a surface in the octant. It is not {\it a 
priori} clear that this can be arranged. It depends on the arrangement of 
the octant picture. As a matter of fact it works nicely if the corners of 
the octant are made to represent separable states (and it does not work nicely 
if the corners represent the Bell basis). Explicitly, the state is 
separable if the rank of $C_{ij}$ is unity. Using the homogeneous coordinates 
introduced already in eq. (\ref{coordinates}) this means that 

\begin{equation} Z^0Z^3 - Z^1Z^2 = 0 \ . \end{equation}

\noindent In our coordinates this corresponds to the two real equations 

\begin{equation} n_0n_3 - n_1n_2 = 0 \label{sep} \end{equation}

\begin{equation} {\nu}_1 + {\nu}_2 - {\nu}_3 = 0 \ . \label{phases} 
\end{equation}

\noindent The equation does separate into two equations, one independent 
of the phases and the other involving only them. This is a picture that can be 
drawn. 

The surface in the 3-torus is, in itself, a flat 2-torus. The surface in the 
octant is easy to draw too because it has an interesting structure. 
In eq. (\ref{separable}), keep the state of the second subsystem fixed. 
Say $b_0/b_1 = ke^{i{\phi}}$, where $k$ is a real number. This implies that 

\begin{equation} \frac{Z^0}{Z^1} = \frac{b_0}{b_1} \hspace{5mm} \Rightarrow 
\hspace{5mm} n_0 = kn_1 \end{equation}

\begin{equation} \frac{Z^2}{Z^3} = \frac{b_0}{b_1} \hspace{5mm} \Rightarrow 
\hspace{5mm} n_2 = kn_3 \ .  \end{equation}

\noindent As we vary the state of the first subsystem we sweep out a curve 
in the octant that is in fact a geodesic (the intersection between the 3-sphere 
and two hyperplanes through the origin in the embedding space). In our picture 
this is a straight line. In this way we see that the surface in the picture is 
a ruled surface swept out by two families of straight lines. If the octant 
of the 3-sphere (that is the interior of our gnomonic tetrahedron) had been flat 
this would have been an intrinsically curved surface, as a glance at Figure 6 
shows. But the true intrinsic geometry of the surface we see in the picture 
is flat. To see this it is convenient to coordinatize the octant by means 
of Euler angles; 

\begin{equation} \left( \begin{array}{c} n_0 \\ n_1 \\ n_2 \\ n_3 \end{array} 
\right) = \left( \begin{array}{c} \sin{\frac{{\tau} - 
{\phi}}{2}}\sin{\frac{\theta}{2}} 
\\ \sin{\frac{{\tau} + {\phi}}{2}}\cos{\frac{\theta}{2}} \\ 
\cos{\frac{{\tau} - {\phi}}{2}}\sin{\frac{\theta}{2}} \\ 
\cos{\frac{{\tau} + {\phi}}{2}}\cos{\frac{\theta}{2}} \end{array} \right) 
\ . \end{equation}

\noindent A straightforward calculation now shows that the separability 
condition (\ref{sep}) is 
equivalent to ${\phi} = 0$. The coordinate ${\tau}$ varies with the state of 
the first subsystem, and the coordinate ${\theta}$ with the other. The intrinsic 
metric of the surface that we see in the octant is 

\begin{equation} ds^2 = \frac{1}{4}(d{\tau}^2 + d{\theta}^2) \ . \end{equation}

\noindent Hence it is an intrinsically flat surface embedded in a curved space; 
in the language of toric geometry the "manifold with corners" of 
${\bf CP}^1\times {\bf CP}^1$ is a flat square.

This little calculation has an interesting interpretation that we mention for 
the benefit of those readers who are familiar with the Hopf fibration of the 
3-sphere \cite{Urbantke}. The 3-sphere can be filled with a congruence of 
nowhere vanishing geodesics ("Villarceau circles") that twist around each 
other but never meet. 
The Euler angles are adapted to this congruence in such a way that ${\tau}$ runs 
along the geodesics and the latter are labelled by ${\theta}$ and ${\phi}$. Thus 
we see that our surface is made up of a one parameter family of "Hopf fibres"; 
actually there is another Hopf fibration with the opposite twist and this 
explains why the surface is ruled by two families of intersecting geodesics. 

We are now ready to look at the space of separable states in ${\bf CP}^3$. 
It is given in Figure 6. 

\begin{figure}
        \centerline{ \hbox{
                \epsfig{figure=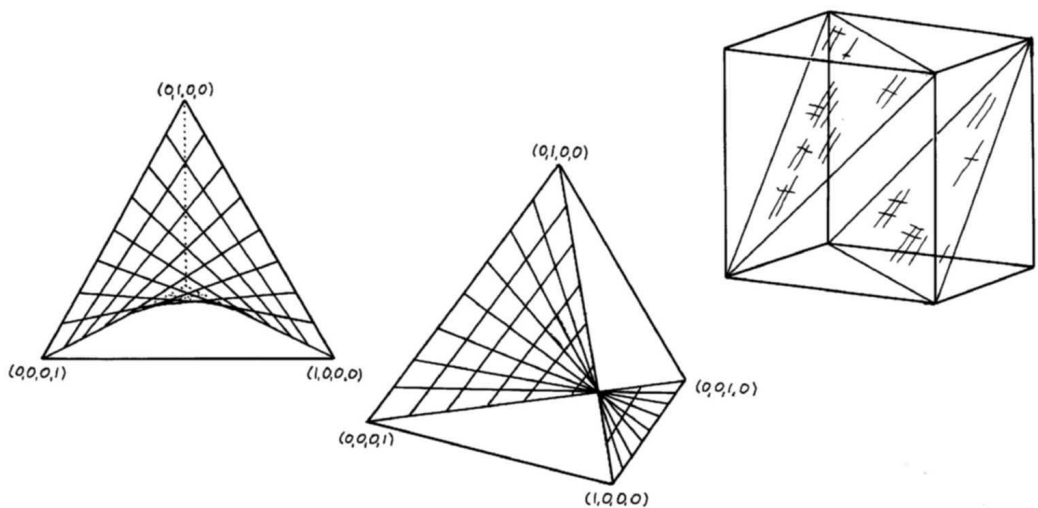,width=14cm}}}
        \caption{{\small The separable states, or the submanifold ${\bf CP}^1 
\times {\bf CP}^1$ in ${\bf CP}^3$. It appears as a surface ruled by geodesics 
in the octant, and it must be remembered that there is a 2-torus sitting in the 
3-torus over each of its interior points. The intrinsic geometry of the 
"manifold with corners", that is the ruled surface that we see in the octant, 
is actually flat. Two different perspectives of the octant are shown. 
The 3-torus is shown schematically as a cube.}}
        \label{fig:sex}
\end{figure}

We now turn to the maximally entangled states. According to eq. (\ref{SU}) 
a state is maximally entangled if and only if the matrix $C_{ij}$ is unitary. 
Since an overall factor of this matrix is irrelevant for the state we can 
use it to adjust the determinant of the matrix to equal one. The state now 
determines the matrix up to multiplication with an $N$th root of unity. 
We arrive at the conclusion that the space of maximally 
entangled states form the group manifold $SU(N)/{\bf Z}^N = U(N)/U(1)$ 
\cite{Fivel} \cite{Werner}. The reason why this space turns up is that, like the 
separable states, the maximally entangled ones forms an orbit of the group 
of local unitary transformations. The group manifold $SU(N)/{\bf Z}^N$ and 
the product space ${\bf CP}^n\times {\bf CP}^n$ then 
spring to mind as the two most obvious candidates.

When $N = 2$ we have $SU(2)/{\bf Z}^2 = SO(3)$, and this happens to be the real 
projective space ${\bf RP}^3$. To see what it looks like in the picture we 
parametrize the unitary matrix $C_{ij}$ as 

\begin{equation} C_{ij} = \left( \begin{array}{cc} {\alpha} & {\beta} \\ 
- {\beta}^* & {\alpha}^* \end{array} \right) \hspace{5mm} \Rightarrow 
\hspace{5mm} Z^{\alpha} = ({\alpha}, {\beta}, - {\beta}^*, {\alpha}^*) 
\ . \end{equation}

\noindent In our coordinates this yields three real equations, namely 

\begin{equation} n_0 = n_3 \hspace{12mm} n_1 = n_2 \end{equation}

\begin{equation} {\nu}_1 + {\nu}_2 - {\nu}_3 = {\pi} \ . \end{equation}

\noindent In the octant this is a single geodesic connecting the entangled 
edges and 
passing through the center of the tetrahedron. In the 3-torus it is a surface 
representing 
a flat 2-torus. So it appears as a one parameter family of 2-tori, in fact 
an ${\bf RP}^3$, as it should. Note that the geodesic in the octant crosses 
the separable surface where the 3-torus achieves its maximum size; this is 
how it manages to keep its distance from the entangled states (namely, every 
maximally entangled state is at the distance ${\pi}/4$ from the separable 
surface). This is shown in Figure 7, where we also show the location of the 
maximally entangled Bell basis  

\begin{equation} |{\psi}^{\pm }\rangle = |0\rangle |1\rangle \pm |1\rangle 
|0\rangle \hspace{8mm} |{\phi}^{\pm }\rangle = |0\rangle |0 \rangle 
\pm |1\rangle |1\rangle \ . \end{equation}

\begin{figure}
        \centerline{ \hbox{
                \epsfig{figure=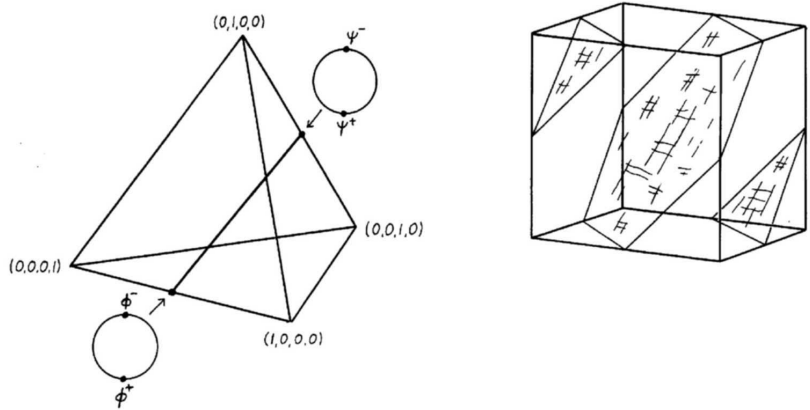,width=11cm}}}
        \caption{{\small The maximally entangled states form an 
${\bf RP}^3$'s worth of points, all at distance ${\pi}/4$ from the closest 
separable states. In the octant we see a single straight line; the 
location of the standard Bell basis is also shown. There is a 2-torus in the 
3-torus (again shown schematically as a cube) over each interior point.}}
        \label{fig:sju}
\end{figure}

\begin{figure}
        \centerline{ \hbox{
                \epsfig{figure=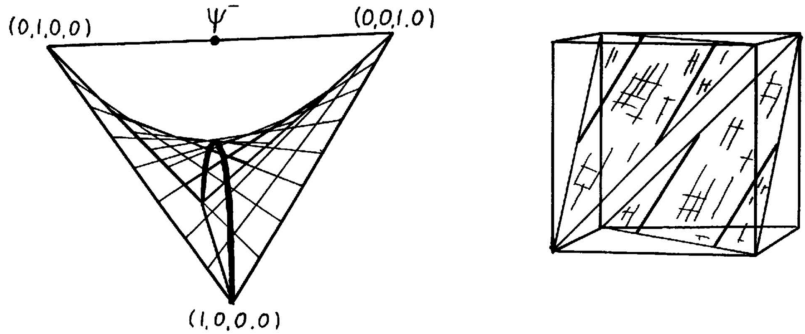,width=11cm}}}
        \caption{{\small If the global system is in the Bell state 
$|{\Psi}^-\rangle $ and if we perform a complete measurement on one of the 
subsystems we end up on a sphere that lies on the separable surface. It is a 
one parameter family of circles as it should be. We show one such circle 
lying on the separable surface in the 3-torus on the right hand side.}}
        \label{fig:atta}
\end{figure}

Finally, let us illustrate the collapse of the wave function. If we perform a 
measurement on one of the subsystems the state of the composite system will 
"jump" in the general direction of the separable surface. If the measurement is 
a von Neumann measurement we end up on the surface of separable states. The 
question is, where? The most likely possibility is on that point on the 
separable surface that is closest to the state we started out from. In the 
generic case this 
point is unique. (To compute it, first use local unitaries to bring the given 
state to the Schmidt form. Unitary transformations are isometries and do not
change distances. It is now easy to see that the closest separable state is in 
fact the nearest corner of the Schmidt simplex. Generically this is unique.) 
Maximally entangled states are exceptional in that there is an 
entire 2-sphere's worth of points on the separable surface at equal distance 
from the given state. This 2-sphere is not a complex projective line because 
its radius 
is $1/\sqrt{2}$ whereas a complex projective line is a sphere with radius $1/2$. 
We are equally likely to land anywhere on this 2-sphere. This is basically 
the statement that if the global state is completely entangled then we do not 
know anything about the state of a subsystem. So Figure 8 picture depicts 
that 2-sphere of separable states that is closest (at distance ${\pi}/4$) 
to the Bell state $|{\Psi}^-\rangle $. It consists of all states of the form 
$|+\rangle_n |-\rangle_n$ for some choice of direction ${\bf n}$. 


\vspace{1cm}

{\bf 4. Submanifolds of fixed entanglement.}

\vspace{5mm}

\noindent We can go on to ask about states with some intermediate 
degree of entanglement. In the particular case when $N = 2$ the remainder 
of ${\bf CP}^{N^2 - 1}$ is foliated by a one-parameter family of five 
dimensional 
orbits of local unitaries. We will not draw such a submanifold here but 
it can actually be done, and the result is much nicer than one might have 
thought. A pure state of intermediate entanglement has the Schmidt decomposition 

\begin{equation} |{\Psi}\rangle = \cos{\sigma}|++\rangle + \sin{\sigma}|--
\rangle \ , \hspace{8mm} 0 < {\sigma} < {\pi}/4 \ . \end{equation} 

\noindent The entanglement grows monotonically with the Schmidt angle 
${\sigma}$. The partially traced density matrix (that arises when we 
trace over one of the subsystems) then has two unequal and 
non-zero eigenvalues ${\lambda}_1 = \cos^2{\sigma}$ and ${\lambda}_2 = 
\sin^2{\sigma}$. A minor calculation verifies 
that the partially traced density matrix has these eigenvalues if and only if 

\begin{equation} \cos{({\nu}_3 - {\nu}_1 - {\nu}_2)} = 
\frac{n_0^2n_3^2 + n_1^2n_2^2 - \cos^2{\sigma}\sin^2{\sigma}}
{2n_0n_1n_2n_3} \ . \label{36} \end{equation}

\noindent When this equation has any solutions at all, it describes a 2-torus 
in the 3-torus, but this time its position within the 3-torus depends on 
where we are in the octant. There are three dimensions left to account for, 
and the pleasant surprise is that this appears as a volume that only 
partly fills the octant. Its boundaries are obtained by setting the 
right hand side of the preceding equation equal to $\pm 1$, that is by 

\begin{equation} (n_0n_3 - n_1n_2)^2 \leq \cos^2{\sigma}\sin^2{\sigma} 
\leq (n_0n_3 + n_1n_2)^2 \ . \end{equation}

\noindent When ${\sigma}$ is small it lies close to the separable surface, 
while 
for ${\sigma}$ close to ${\pi}/4$ what we see in the octant is a kind of 
three dimensional tube surrounding the maximally 
entangled line. We leave its precise appearance as an exercise for the reader. 
Note that the case $N = 2$ is exceptional; when $N > 2$ the orbits of local 
unitary transformations have codimension larger than one, there is no 
obviously canonical measure of pure state entanglement, and a much more 
intricate picture emerges. For illustrations of the $N = 3$ case consult 
\cite{BZ}.

The situation is qualitatively similar to that of the orbits of $SU(2)$ in 
${\bf CP}^2$ \cite{Nuno}, also 
in the sense that the octant picture of the orbit looks nice but is poorly 
suited to do calculations. Once it is understood what a $U(2)\times U(2)$ orbit 
looks like we can watch it grow and shrink as we increase the entanglement. 
But to understand its intrinsic geometry it is better to proceed as follows: 
Choose a point in the orbit given in Schmidt form by 

\begin{equation} C^{S}_{ij} = \left( \begin{array}{cc} \cos{\sigma} 
& 0 \\ 0 & \sin{\sigma} \end{array}\right) \ , \label{39} \end{equation}

\noindent where $C_{ij}$ is the matrix introduced in eq. (\ref{17}) and 
${\sigma}$ runs from $0$ to ${\pi}/4$; ${\sigma}$ is the Schmidt angle and 
increases monotonically with the entanglement. An arbitrary point in the 
orbit can be reached from the given one by means of unitary transformations 
of the factor Hilbert spaces. Actually $SU(2)$ transformations are enough, 
and these can be parametrized with Euler angles. That is to say that an 
arbitary point in the orbit can be given by a matrix $C_{ij}$ defined by 

\begin{equation} C = e^{-i{\phi}_1L_z}e^{i{\theta}_1L_y}e^{-i{\tau}_1L_z}C^S
e^{-i{\tau}_2L_z}e^{- i{\theta}_2L_y}e^{-i{\phi}_2L_z} \label{40} \end{equation}

\noindent where $L_y$ and $L_z$ are the angular momentum operators in the 
standard representation (and one of the group elements appears transposed in 
the formula). 
In the calculation one sees that only ${\tau} \equiv {\tau}_1 + {\tau}_2$ 
matters. After a still fairly elaborate calculation we obtain the Fubini-Study 
metric in the form 

\begin{equation} ds^2 = d{\sigma}^2 + dl^2 \ , \end{equation}

\noindent where $dl^2$ is the intrinsic metric on the five dimensional orbit 
labelled by $0 < {\sigma} < {\pi}/4$. Explicitly 

\begin{eqnarray} dl^2 = \frac{1}{4}( d{\theta}^2_1 + \sin^2{{\theta}_1}
d{\phi}_1^2 + d{\theta}^2_2 + \sin^2{{\theta}_2}d{\phi}_2^2 + \nonumber \\ 
\ \nonumber \\
+ 2\sin{2{\sigma}}(\cos{\tau}\sin{{\theta}_1}\sin{{\theta}_2}d{\phi}_1d{\phi}_2 
- d{\theta}_1d{\theta}_2 - \nonumber \\ 
\ \\
- \sin{\tau}\sin{{\theta}_2}d{\theta}_1d{\phi}_2 - 
\sin{\tau}\sin{{\theta}_1}d{\theta}_2d{\phi}_1) \nonumber \\
\ \nonumber \\ 
+ \sin^2{2{\sigma}}(d{\tau} + \cos{{\theta}_1}d{\phi}_1 + 
\cos{{\theta}_2}d{\phi}_2)^2) \nonumber \end{eqnarray}

\noindent where $0 < {\phi}_1, {\phi}_2 , {\tau} < 2{\pi}$, $0 < {\theta}_1, 
{\theta}_2 < {\pi}$. When ${\sigma} = 0$ this reduces to the metric of 
${\bf S}^2\times {\bf S}^2$ as it should, while the coordinates misbehave when 
${\sigma} = {\pi}/4$. The square root of the determinant of this metric is 

\begin{equation} \sqrt{g} = \frac{1}{2^5}\cos^2{2{\sigma}}\sin{2{\sigma}}
\sin{{\theta}_1}\sin{{\theta}_2} \ . \end{equation} 

\noindent (Actually the clever way to compute this is to go via the symplectic 
form presented in section 6. The coordinates used here are well adapted for 
this task.) The volume of a given orbit can now be computed 
and is found to be 

\begin{equation} \mbox{vol}({\sigma}) = {\pi}^3\cos^2{2{\sigma}}\sin{2{\sigma}}
\ . \end{equation}

\noindent Dividing by the volume ${\pi}^3/6$ of ${\bf CP}^3$ we obtain a 
probability distribution $P({\sigma})$ for the Schmidt angle ${\sigma}$. 
Note that the unitarily invariant distribution over the set of pure states 
used here induces the unitary distribution inside the Bloch ball for the 
mixed states ${\rho}_A = \mbox{Tr}_B|{\psi}\rangle \langle {\psi}|$. 
\cite{Sommers}. 

Another coordinatization of the orbit works better when the entanglement is 
close to maximal. Above we used two arbitrary unitary matrices $u_1$ and 
$u_2$ to write 

\begin{equation} C = u_1C^Su_2 \ . \end{equation} 

\noindent If we introduce $u_3 = u_1u_2$ this becomes

\begin{equation} C = u_1C^Su_1^{-1}u_3 \ . \end{equation}

\noindent Now one Euler angle in $u_1$ is irrelevant; moreover we see directly 
that in the maximally entangled case---when $C^S$ becomes diagonal---the orbit 
collapses to the group manifold of $SU(2)/{\bf Z}^2 = {\bf RP}^3$. One can 
check that the embedding is isometric.

Since we have a foliation of ${\bf CP}^3$ with five dimensional 
hypersurfaces---except for the exceptional subsets where the entanglement either 
vanishes or is maximal---it is interesting to look at the second fundamental 
form of these hypersurfaces. One finds that the trace of the extrinsic 
curvature tensor is 

\begin{equation} K = \frac{4}{\cos{2{\sigma}}\sin{2{\sigma}}}(\cos^2{2{\sigma}} 
- 
2\sin^2{2{\sigma}}) \ . \end{equation}

\noindent A foliation with the property that $K$ is constant on each 
hypersurface 
is called a constant mean curvature foliation. We observe that $K = 0$ when 
$\tan{2{\sigma}} = 1/\sqrt{2}$, that is when the volume of the orbit is maximal. 

Let us emphasize again that this state of affairs is special to two qubit 
entanglement; in higher dimensions a much more intricate story emerges. 
See ref. \cite{Sino} for the dimensions of the local orbits that appear in 
the $N \times N$ case. 


\vspace{1cm}

{\bf 5. An Aside on the Statistical Geometry of Simplices.}

\vspace{5mm}

\noindent Due to obvious limitations we cannot literally draw our picture 
when the dimension of the Hilbert space exceeds four. There is one particular 
aspect of entanglement that we can draw up to dimension $4 \times 4$, 
however. This is the Schmidt simplex, that is to say all states that 
have the form given in eq. (\ref{Schmidt}). As we mentioned a central 
fact is that given any state there is a local unitary transformation that 
brings it into the Schmidt simplex. It is a simplex essentially because of 
eq. (\ref{Scmidtsimplex}). It is clear that the Schmidt simplex is an $N-1$ 
dimensional (hyper-)face of the hyperoctant that forms a part of our picture. 
It is easy to check that its intrinsic geometry---induced by the Fubini-Study 
metric---is round, so in itself it forms the (hyper-)octant of a round sphere. 
Using gnomonic coordinates we draw it as a flat simplex. 

The Schmidt simplex should not be confused 
with another simplex having the same pure states sitting in its corners, 
namely the statistical simplex whose points represent density matrices in 
the convex cover of these $N$ states. The latter is an $N - 1$ dimensional 
subset of the $N^2 - 1$ dimensional space of density matrices, consisting 
of density matrices that can be simultaneously diagonalized. Usually 
the probabilities $p_i$ are used as barycentric coordinates on this simplex, 
which then consists of all density matrices of the form

\begin{equation} {\rho} = \sum_{i = 0}^n p_i|i\rangle \langle i| \ , 
\hspace{5mm} \sum_{i = 0}^n p_i = 1 \ . \end{equation}

\noindent There is an obvious flat metric on this simplex that has the 
property that the density matrices obtained by taking the mixture of two 
density matrices in the set appears as a straight line connecting two 
extreme points representing the original pair of density matrices. There is 
also a round metric on this simplex that is induced by the Bures metric on the 
set of all density matrices \cite{Uhlmann}; it 
captures the statistical geometry of density matrices \cite{Caves}. 

A possible confusion now arises because we have one round and two flat 
metrics on the same simplex, the latter two being the metric that makes 
statistical mixtures appear as straight lines and the metric that 
naturally exists on the gnomonic coordinate plane. These are not the 
same. If we draw the round simplex using gnomonic coordinates the 
statistical mixtures will be represented by curves, as in Figure 9. This 
must be kept in mind if we ask (say) how far away the corners are from 
the interior: The usual statement \cite{Wootters} that the corners are 
further away than they seem refers to the coordinates $p_i$; the 
significance of these coordinates is that they manifest the convexity 
properties of the simplex.  

\begin{figure}
        \centerline{ \hbox{
                \epsfig{figure=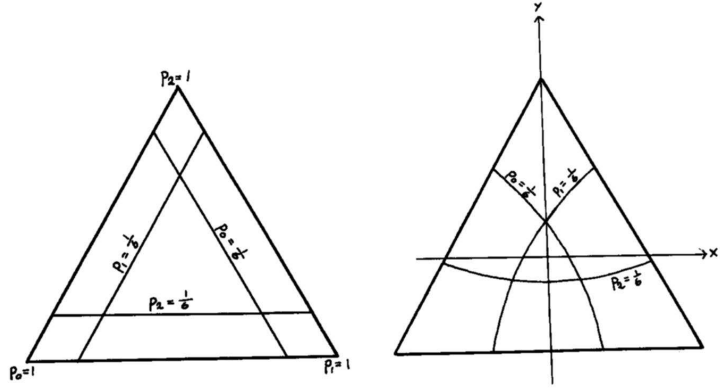,width=10cm}}}
        \caption{{\small The statistical simplex of diagonal density matrices 
of size $N = 3$, drawn with the eigenvalues $p_i$ as barycentric coordinates 
(left), and the same simplex drawn as a round simplex using gnomonic 
coordinates $x, y$ (right). 
Corners represent pure states while the centers of the triangles represent 
the maximally mixed state. The round (statistical) distance 
from a given point to the nearest corner appears too short in the former 
picture and too long in the latter.}}
        \label{fig:nio}
\end{figure}

Note also that we have two different round simplices with pure 
states in its corners, namely the Schmidt simplex that consists of pure 
states, and the statistical simplex that consists of density matrices.


\vspace{1cm}

{\bf 6. An aside on symplectic geometry.}

\vspace{5mm}

\noindent The manifold ${\bf CP}^n$ does not enjoy a metric geometry only. 
It also has a 
symplectic structure closely allied to it; in physicist's language it is a 
phase space equipped with a Poisson bracket in a natural way. In more 
mathematical 
language there is a symplectic (closed, non-degenerate) 2-form around. Since 
we are praising the virtues of a special coordinate system here it seems 
natural to mention that the symplectic 2-form ${\Omega}$ also takes a simple 
form in this coordinate system. In effect 

\begin{eqnarray} {\Omega} = 2i\frac{Z\cdot \bar{Z} dZ\cdot \wedge d\bar{Z} 
- \bar{Z}\cdot dZ \wedge d\bar{Z}\cdot Z}
{Z\cdot \bar{Z} Z\cdot \bar{Z}} = \hspace{6mm} \nonumber \\
\ \\
\hspace{6mm} = 4(n_1dn_1\wedge d{\nu}_1 + n_2dn_2\wedge d{\nu}_2 + 
n_3dn_3\wedge d{\nu}_3) \ , \nonumber \end{eqnarray}

\noindent where the last line is for $n = 3$. Hence we can think of $n_i^2$ as 
being "canonically conjugate" to the phase ${\nu}_i$; in effect these are 
action-angle variables. There is a nice interplay 
between the symplectic geometry and the geometry of entanglement. In particular 
a straightforward calculation verifies that the space of maximally entangled 
states is a {\it Lagrangian submanifold}. A submanifold of dimension $D$ in a 
symplectic space of dimension $2D$ is said to be Lagrangian if and only if the 
symplectic form induced on the submanifold 
by the embedding vanishes; mathematicians know that $SU(N)/{\bf Z}^N$ is a 
Lagrangian submanifold of ${\bf CP}^{N^2 - 1}$ but no complete classification 
of Lagrangian submanifolds is available \cite{Bryant}.

Let us now specialize to ${\bf CP}^3$, and give the symplectic form in the 
coordinates introduced in eqs. (\ref{39})-(\ref{40}): 

\begin{eqnarray} {\Omega} = 2\sin{2{\sigma}}(d{\sigma}\wedge d{\tau} + 
\cos{{\theta}_1}d{\sigma}\wedge d{\phi}_1 + 
\cos{{\theta}_2}d{\sigma}\wedge d{\phi}_2) + \nonumber \\ 
\ \\
+ \cos{2{\sigma}}(\sin{{\theta}_1}d{\theta}_1\wedge 
d{\phi}_1 + \sin{{\theta}_2}d{\theta}_2\wedge d{\phi}_2) \ . \nonumber 
\end{eqnarray}

\noindent On the five dimensional orbits of local unitaries the first line goes 
away since ${\sigma}$ is constant. The symplectic form is then degenerate and 
goes smoothly over to the symplectic form on ${\bf S}^2\times {\bf S}^2$ 
as ${\sigma}$ 
goes to zero. In the language often used by physicists interested in constrained 
systems \cite{Ashtekar} the equation (\ref{36}) that defines the orbit is a 
first class constraint and the coordinate ${\tau}$ runs along the gauge orbits. 

We observe that the volume element discussed in section 5 is easy to derive 
using 

\begin{equation} \sqrt{g}d^6x = \frac{1}{3!}(\frac{1}{4}{\Omega})\wedge 
(\frac{1}{4}{\Omega})\wedge (\frac{1}{4}{\Omega}) 
\ . \end{equation}

\noindent This alternative way of computing the volume element is always open 
on K\"ahler manifolds such as ${\bf CP}^n$.


\vspace{1cm}

{\bf 7. Envoi.}

\vspace{5mm}

\noindent We hope that our results serve to bring home the fact that 
${\bf CP}^n$ has an interesting geometry, that this geometry can be 
visualized without too much effort, and that the resulting picture has 
something to tell us about the physics 
of entanglement---at the very least, that it serves to illustrate it.

\vspace{1cm}

{\bf Acknowledgement:}

\vspace{5mm}

\noindent We thank Torsten Ekedahl for some help.

\end{document}